\documentclass[12pt]{iopart}
\usepackage{iopams}  
\usepackage{color}
\usepackage{graphicx}
\usepackage{lmodern}
\begin{document}

\title{Electron-phonon interaction in a local region}

\author{K.V. Reich}

\address{Ioffe Physical-Technical Institute of the Russian Academy of Sciences}
\ead{Reich@mail.ioffe.ru}
\begin{abstract}
The paper reports on a study of electron-phonon interaction within a limited nanosized region. We invoked the modified Fr\"{o}hlich's Hamiltonian to calculate the electron self-energy, as well as the elastic and inelastic scattering cross sections. New effects have been revealed, more specifically: a bound state forms within the limited nanosized region, electrons undergo resonant elastic scattering, with strong inelastic scattering being possible from this state even at low electron energies. The effect of scattering on the magnetic-field-independent dephasing time, in particular, in a diamond-decorated carbon nanotube, has been determined. The effect of strong inelastic electron scattering on thermal resistance at the metal-insulator interface is discussed.
\end{abstract}

\maketitle

\section{Introduction}
Interaction of electrons with vibrational modes, phonons, in nanosized regions has been attracting recently considerable interest \cite{Galperin}. This interaction underlies many mesoscopic phenomena; to cite an example, it was found to affect strongly the current flowing through a nanosized region \cite{Reich2}, and it is this interaction that mediates the electron dephasing time \cite{Bauerle}. It is well known that in the low-temperature domain the Kondo effect contributes significantly to the magnitude of the dephasing time. This time depends substantially on magnetic field \cite{Micklitz}. It was predicted recently \cite{Dora} and demonstrated subsequently experimentally that the vibrational mode affects the dephasing time of electrons propagating in an amorphous film \cite{expCu} or a graphene sheet \cite{Giant}.  Interestingly, this dephasing time can be quite large, and it does not depend on magnetic field. Electron-phonon interaction mediates transport phenomena in nanostructures as well. We are witnessing presently intense research interest in composite materials based on the metal-diamond pair \cite{Kidalov}. Thermal conductivity of such composites depends substantially on the possibility of energy transfer between the electron gas in the metal and the diamond phonons. This process occurring in the vicinity of the metal-diamond boundary, it appears essential that the theory of this phenomenon takes electron-phonon interaction in the nanosized region into proper account.

All studies in the theory of electron-phonon interaction consider either spatially-uniform interaction \cite{unit} or interaction at a point \cite{Dora}. In both cases, the approach used invokes either the Fr\"{o}hlich or the Holstein Hamiltonian. It has recently been demonstrated that inclusion of the spatial distribution of local vibrational modes affects noticeably the properties of the system under study \cite{Inelasticexp}. The author is, however, unaware of any publication which would consider the dependence of the properties of a system on the size of the region supporting electron-phonon interaction. This suggests the need of tackling the fundamental problem of extending the results amassed on point and spatially uniform electron-phonon interaction to cover the case of interaction within a local region of finite dimensions.

The goal of the present study is development of a theory of electron transport through a nanosized region with local electron-phonon interaction. The study is conducted for the particular case of the impact of electron-phonon interaction in a local region on current flow along a carbon nanotube decorated by nanodiamonds, a novel material obtained recently \cite{Reich1}. It is maintained \cite{emission} that such interaction affects markedly the electronic properties of this material. We are also going to consider the effect of strong inelastic scattering on thermal resistance at the metal-insulator boundary \cite{Kidalov}.

\section{Model}
\subsection{Hamiltonian and the electron self-energy}
We start by constructing a Hamiltonian for systems with local electron-phonon interaction. As the starting point will serve the Fr\"{o}hlich Hamiltonian into which we shall introduce a coordinate-dependent function describing electron-phonon interaction. The Hamiltonian of the system can be written in the form $H=H_0+H_{ep}$. As usual,
\begin{equation}
 H_0 = \sum_{\bi{p}} \varepsilon_{\bi{p}} c_{\bi{p}}^{+} c_{\bi{p}} + \sum_{\bi{q}} \hbar \omega_{\bi{q}} a_{\bi{q}}^{+} a_{\bi{q}}
\end{equation}
- is the Hamiltonian of noninteracting electrons and phonons, and
\begin{equation}
\label{Hep}
 H_{ep}=\int d \bi{r} \rho(\bi{r}) V(\bi{r})
\end{equation}
- is the electron-phonon interaction Hamiltonian. The Hamiltonians are presented in their standard notation. We used here operators of creation and destruction of electrons, $c^{+}_{\bi{p}}, c_{\bi{p}}$ , and phonons, $a^{+}_{\bi{q}}, a_{\bi{q}}$.  The energies of the electron, $\varepsilon_{\bi{p}}$, and of the phonon,  $\hbar \omega_{\bi{q}}$ , depend accordingly, on the momentum ${\bi{p}}$ and the wave vector ${\bi{q}}$. We shall assume in what follows that Planck's constant $\hbar=1$. In the $H_{ep}$ Hamiltonian, $\rho(\bi{r})$ is the electron density operator, and $V(\bi{r})$, operator of the electron-phonon interaction potential. 

By local electron-phonon interaction we are going to understand here the interaction which in a certain specified region is stronger than in the surrounding regions. We will assume that electron-phonon interaction exists in such a region only. To discriminate this region and identify properly its differences from the other parts of the sample, introduce the localization function $\lambda(\bi{r})$. This function is of order unity within the region of interaction, and small outside it. This may be, for instance, a Gaussian function. We shall write the interaction potential in the form of a product $\lambda(\bi{r}) V(\bi{r})$ rather than in its conventional notation $V(\bi{r})$. This form of the potential underlies all of our subsequent consideration. We further write Hamiltonian  (\ref{Hep}) in terms of the Fourier transforms of the factors it contains. Integration over the coordinate $\bi{r}$ yields:
\begin{equation}
 H_{ep}= \sum_{\bi{k}} \lambda(\bi{k}) \sum_{\bi{q}} \rho(\bi{q}+\bi{k}) V(\bi{q})
\end{equation}
We next express the quantity $\rho(\bi{q}+\bi{k}) V(\bi{q})$ in its standard form through the operators of creation and destruction of electrons and phonons \cite{Mahan}. Thus we have
\begin{equation}
\label{HepEnd}
H_{ep}=\sum \limits_{\bi{k}} \lambda (\bi{k}) \left[ \sum \limits_{\bi{p},\bi{q}} M_{\bi{q}} c^{+}_{\bi{p}+\bi{q}+\bi{k}}c_{\bi{p}} (a_{\bi{q}}+a^{+}_{-\bi{q}}) \right]
\end{equation}
This expression contains the coefficient of interaction of electrons with phonons $M_{\bi{q}}$. We readily see that for the localization function $\lambda(\bi{r})=1$ Hamiltonian (\ref{HepEnd})  transforms throughout the sample to the Hamiltonian of spatially uniform electron-phonon interaction \cite{unit}, while for  $\lambda(\bi{r})=\delta(\bi{r})$ it becomes the point-interaction Hamiltonian \cite{Dora}.
We shall formulate now the electron scattering problem within a region with local electron-phonon interaction. We shall consider Hamiltonian  $H_{ep}$ as a perturbation and perform the calculations as this is done in Ref. \cite{Mahan}. Green's function $G(\bi{p},\varepsilon)$ written in the momentum representation for the complete Hamiltonian $H$ is related to Green's function $G(\bi{p},\varepsilon)$ corresponding to the unperturbed Hamiltonian $H_0$. This relation can be presented in the form $G=G_0+G_0 \Sigma G_0$ , where $\Sigma(\bi{p},\varepsilon)$ - is the electron self-energy (SE). The procedure of SE calculation is known to consist essentially in calculating the interaction Hamiltonian (\ref{HepEnd}) averaged over the electron and phonon states.  An important note is here in order; in view of our subsequent calculations of the cross sections of scattering from a region with local electron-phonon interaction, one can conveniently consider, from a formal standpoint, a system combining many regions with local-phonon interaction whose concentration is $n = 1$. In this case, one will have to perform averaging over all positions of such regions, an approach adopted in \cite{Mahan}. To do this, we transform Hamiltonian  $H_{ep}$ to operators in the interaction representation. The localization function $\lambda(\bi{r})$ being time-independent, the interaction Hamiltonian contains the dependence on time only in the bracketed factor in Eq.  (\ref{HepEnd}). Therefore, the procedure of SE calculation for local electron-phonon interaction is similar to that for electron-phonon interaction in the spatially uniform case. In this Section, we shall restrict ourselves to calculating SE in the first approximation. The exact expression for the SE will be developed in the next Section. Thus we have:
\begin{equation}
\label{Xsi}
\Sigma (\bi{p},\varepsilon) = \int \lambda (\bi{p}-\bi{k}) \lambda (\bi{k}-\bi{p}) \Sigma_u (\bi{k},\varepsilon) d \bi{k}
\end{equation}
where, $\Sigma_u(\bi{p},\varepsilon)$ is the self-energy of the electron for the case of spatially uniform electron-phonon interaction.  In order to calculate the SE of electron-phonon interaction in a local region with the help of Eq. (\ref{Xsi}), one will have to know the form this expression assumes in the spatially uniform case,  $\Sigma_u(\bi{p},\varepsilon)$ , which, in its turn, depends on the system being considered. 

Indeed, in the  particular case of Einstein phonons of frequency $\omega_0$:, we have $\Im \Sigma_u(\bi{p},\varepsilon) \sim \Theta (\varepsilon - \omega_0)$, where $\Theta (\varepsilon)$  is the Heaviside step function. We immediately see that the imaginary part of SE depends on energy while being independent of  momentum. Substituting this expression in Eq. (\ref{Xsi}), we see that because integration is performed over the momentum $\bi{k}$, the imaginary part of the SE in the case of Einstein phonons will be likewise a stepped function of energy. Analytic continuation of this relation yields, just as demonstrated in \cite{gogolin}, a log dependence of the real part on energy. In other words, consideration of Einstein phonons in a local region does not introduce anything new of particular significance. New effects can be obtained by analyzing the behavior of phonons with a linear dispersion law. This is exactly the case we are going to consider now.

\subsection{\label{sec:effpotential} Effective potential in local electron-phonon interaction}
For free electrons, the real part $\Re {\Sigma_u(\bi{p},\varepsilon)}$  is constant and does not depend on the electron momentum and energy. In other cases, the real part $\Re {\Sigma_u(\bi{p},\varepsilon)}$ does depend on these quantities. To isolate qualitative effects, we shall assume that in the case under consideration the real part $\Re {\Sigma_u(\bi{p},\varepsilon)}$ is constant too, and that it is equal to $\Sigma_0$. Substituting this quantity in Eq. (\ref{Xsi}), we see immediately that $\Re {\Sigma(\bi{p},\varepsilon)}$ is likewise a constant. We are going to use this result in the subsequent Sections. However  to better understand the physical aspects underlying this phenomenon, it will be appropriate to recover here Hamiltonian $H$ and subject it to Bogolyubov's transformation. 
To do this, we rewrite the initial Hamiltonian by changing in (\ref{HepEnd}) the dummy summation variable $\bi{k}$ to $\bi{k}-\bi{p}-\bi{q}$,  and, subsequently, by transforming operators $a_{\bi{q}}$ to the new set 
$$a_{\bi{q}}-\lambda(\bi{q}) \frac{M_{\bi{q}}}{\omega_{\bi{q}}} \sum \limits_{\bi{k}} c^{+}_{\bi{k}} c_{\bi{k}}$$
The sum in this expression, as is well known, is the number of electrons. Since the carbon nanostructures such as nanotubes and graphene can be a semimetal or a semiconductor \cite{semimetal} the low temperature free electron concentration is small. Thus, it is possible use single-electron approximation. On performing the transformation, we see that the Hamiltonian (\ref{HepEnd}) of the system can be presented in the form
\begin{eqnarray}
\fl H= \sum _{\bi{p}} \varepsilon_{\bi{p}} c_{\bi{p}}^{+} c_{\bi{p}} - 2 \sum \limits_{\bi{k} \bi{p} \bi{q}} \lambda(\bi{q}) \lambda(\bi{k}-\bi{q}) \frac{M^2_{\bi{q}}}{\omega_{\bi{q}}} c^{+}_{\bi{k}+\bi{p}} c_{\bi{p}} + \nonumber\\
+ \sum _{\bi{q}} \hbar \omega_{\bi{q}} a_{\bi{q}}^{+} a_{\bi{q}} + \sum \limits_{\bi{q}} \lambda(\bi{q})^2 \frac{M^2_{\bi{q}}}{\omega_{\bi{q}}} +\sum \limits_{\bi{k} \neq \bi{p}, \bi{q}} \lambda(\bi{k}-\bi{p}-\bi{q}) M_{\bi{q}} c^{+}_{\bi{k}} c_{\bi{p}} (a_{\bi{q}}+a^{+}_{-\bi{q}}) \nonumber
\end{eqnarray}

We shall neglect in what follows the term before the last. This is the shift of the origin of energy. The first two terms will be considered as the ''unperturbed'' Hamiltonian for an electron in the region of local electron-phonon interaction. This Hamiltonian can be recast in the form of the Hamiltonian for the electron residing in the effective potential $U(\bi{r})$. In the momentum representation this effective potential assumes the form:
\begin{equation}
\label{effpotential}
U(\bi{k})= - 2 \sum \limits_{\bi{q}} \lambda(\bi{q}) \lambda(\bi{k}-\bi{q}) \frac{M^2_{\bi{q}}}{\omega_{\bi{q}}}
\end{equation}
Rewriting now Eq. (\ref{effpotential}) in the coordinate representation, we can say that the electron resides in a potential well with a characteristic depth $\Sigma_0=\sum_{\bi{q}} M^2_{\bi{q}} / \omega_{\bi{q}}$. A potential similar to the effective potential $U(\bi{r})$ is used in the theory of electron scattering from an impurity. The specific form of potential (\ref{effpotential})  is determined naturally by the structure of the material under study, which is set by the localization function $\lambda(\bi{r})$ and the dispersion dependences of the frequency $\omega_{\bi{q}}$ and of the electron-phonon interaction constant $M_{\bi{q}}$ on the phonon wave vector $\bi{q}$. Nevertheless, one can draw certain qualitative conclusions of a general nature on the behavior of an electron in a potential generated by local electron-phonon interaction. 

We assume the localization function to have the form $\lambda(\bi{r})=exp(-r^2/a^2)$. Our subsequent consideration depends substantially on the size of the system under study. Since the theory developed below is to be applied to carbon nanotubes decorated by diamonds, we are going to study quantitatively here the one-dimensional case. The three-dimensional case will be treated only qualitatively. It is known that for an electron in a potential well there always exists a bound state, both in the one- and two-dimensional case. We are going to denote by $\kappa= \sqrt { 2m |\varepsilon_0| / \hbar^2 }$ the wave vector corresponding to the bound state with energy $\varepsilon_0$. For the characteristic energy scales of the energy $\Sigma_0$ corresponding to the polaron binding energy, and the  interaction region size $a$ of interest to us here, we obtain a shallow potential well with the energy of the level $\varepsilon_0=m a^2 \Sigma_0^2 / 2 \hbar^2$. Equation (\ref{effpotential}) suggests that the effective potential $U(\bi{r}, a)$  tends exponentially to zero for both small and large sizes $a$. Whence it follows that one has to take into account the decrease of the well depth itself. We assume that the wave vector $\kappa$ scales with the size $a$ as  $\kappa= m a \Sigma_0 / \hbar ^2 \exp(-a/a_c)$ . Here the parameter  $a_c$  is the size at which $\kappa$ reaches its maximum value, and the level lies closest to the well bottom.
For qualitative studies of the three-dimensional case, a refinement is in order here, namely, that the bound state in 3D exists only if the characteristic size of the well is larger than its critical value  $\sqrt {\hbar^2 / 2m \Sigma_0}$.

We see that, while proceeding along the lines of perturbation theory, we used the free-electron model (\ref{HepEnd}) as the starting point, we came already in the first approximation to the conclusion that the system should support under certain conditions formation of a bound electron state in a region with local electron-phonon interaction. As we are going to see in Section \ref{sec:scattering}, the existence of such a bound state will produce a considerable impact on the amplitude of electron scattering from a region with local electron-phonon interaction. In any case, this "phonon" bound state resembles in many respects the  bound state in an impurity. The imaginary parts of the SE in systems with impurities and with local interaction are essentially different.

\subsection{\label{sec:imagSE}Imaginary part of the self-energy}
Consider separately calculation of the imaginary part of the SE, $\Im {\Sigma(\bi{p})}$ which is intimately connected with the possibility of emission of a phonon by an electron. To do this, we have first to specify the form of the imaginary part of the SE, $\Im {\Sigma_u(\bi{p},\varepsilon)}$, for an electron in the spatially uniform case. By the theory of unitarity, it can be related to the imaginary parts of Green's functions of the electron, $\Im {G_0}$, and the phonon, $\Im {D_0}$ 
\begin{equation}
\label{Im}
\Im{\Sigma_u (\bi{k},\varepsilon)}=\int M_{\bi{q}}^2\Im{G_0(\bi{k}-\bi{q},\varepsilon-\omega)}) \Im{D_0(\bi{q},\omega)} \frac{d \omega}{2 \pi} d \bi{q} 
\end{equation}
We assume here for the sake of simplicity that the electron-phonon coupling coefficient $M_{\bi{q}}$ does not depend on $\bi{q}$,  $M=\bar{M_{\bi{q}}}$. Although for the acoustic phonons being considered here, $M_{\bi{q}}$ does depend linearly on $\bi{q}$, this standard simplification \cite{electronphonon} can be done by invoking the first mean value theorem for the integral (\ref{Im}). 

For spatially uniform electron-phonon interaction, the calculation is performed assuming $\bi{k} = \bi{p}$ and $\varepsilon=p^2/2m$. For local interaction, these conditions are not met. We have therefore to calculate  $\Im {\Sigma_u(\bi{p},\varepsilon)}$ without making use of the relation between  $\varepsilon$ and $\bi{k}$. Introduce into this equation the expressions for Green's function for free electrons and Green's function for free acoustic phonons having a sound velocity $c$. We finally come to 
\begin{equation}
\label{ImSigma}
\Im{\Sigma_u(k,\varepsilon)}= - \frac{1}{8} \frac{M^2 m}{k} \int \limits_L q dq
\end{equation}
where $L$ - is the region of integration over the phonon wave vector $q$, satisfying the inequality:
\begin{equation}
\label{condq}
\frac{(k-q)^2}{2m} < \varepsilon - cq < \frac{(k+q)^2}{2m}
\end{equation}
In the case of spatially uniform electron-phonon interaction, condition  (\ref{condq}) transforms to the well-known condition that an electron can emit a phonon only for $p > mc$. This threshold condition specifies the possibility of phonon emission by an electron.
For a spatially nonuniform electron-phonon interaction, this inequality has always a solution for $k < \sqrt{2m \varepsilon} = p$. Examining Eq. (\ref{ImSigma}), we see that when inequality (\ref{condq}) is satisfied, the imaginary part of SE  (\ref{Xsi}) becomes nonzero. Thus, in the local electron-phonon interaction region there will always be energy losses involved in the emission of acoustic phonons by an electron. This is the main difference of interaction in a local region from the case of spatially uniform electron-phonon interaction. To conduct a qualitative analysis of the 3D case, we assume that for $k < p$ integration in Eq.  (\ref{ImSigma}) runs over all phonon wave vectors $q$, from zero to the Debye wave vector $k_D$. Substitute Eq. (\ref{ImSigma}) into (\ref{Xsi}). Integration performed in the $pa \to 0$ approximation yields
\begin{equation}
\label{endIm}
\Im{\Sigma(\varepsilon,p)}= - \Sigma_0 (pa)^2.
\end{equation}
Thus, in the case of local electron-phonon interaction, the imaginary part of the SE will be proportional to the electron energy and does not have a threshold. The dependence of the SE, Eq.  (\ref{Xsi}) , on momentum and size of the electron-phonon interaction region in the 1D case is presented in graphical form in Fig. 1. Apart from this, numerical calculations revealed that for each value of the electron momentum before interaction there is an optimum size for which the probability of phonon emission is maximal  $a_{max}$. The smaller the electron momentum, the larger is this region.

\begin{figure}
\includegraphics[width=0.9\linewidth]{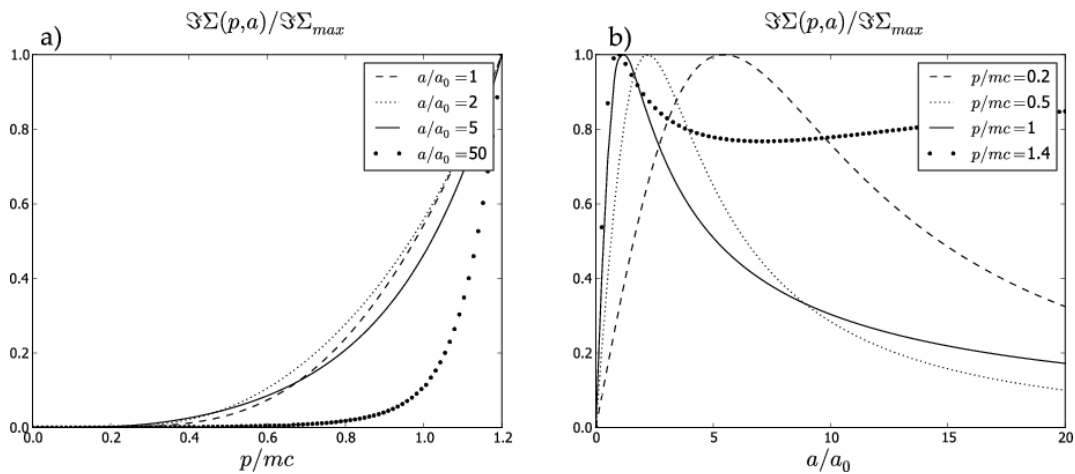}
\caption{
(a) The imaginary part of the SE, $\Im{\Sigma}$ plotted for the 1D case vs. momentum $p$ (in $mc$ units) for different sizes of the local region expressed in  units of  $a_0= 1/mc$.  In electron-phonon interaction localization regions of larger sizes, the dependence acquires a threshold pattern. As the size of such a region decreases, the threshold disappears, i.e., the probability of phonon emission is no longer zero. (b) The dependence of  $\Im{\Sigma}$ on size $a$ for different momenta. For momenta less than $mc$, there exists an optimal size of the electron-phonon interaction localization region, for which the probability to emit a phonon is maximal. For momenta above $mc$,  the size of the region is no longer essential. To present all the graphs on the same scale, they are normalized against their maximum values $\Im{\Sigma_ {max}}$ within a given argument interval.}
\label{fig1}
\end{figure}

\section{\label{sec:scattering}  Scattering amplitude}

In the preceding Section we have considered the SE for an electron interacting with phonons in a local nanosized region in the first approximation. Let us calculate now the values of the SE, $\Sigma(\bi{p},\varepsilon)$, with inclusion of the other terms of the series. We shall neglect, on the one hand, the effect of coherent phonon emission, and on the other, coherent scattering from two regions with electron-phonon interaction. This makes the limiting transition to consideration of a system with spatially uniform electron-phonon interaction impossible. It can readily be shown that  $\Sigma(\bi{p},\varepsilon) = F(\bi{p},\bi{p},\varepsilon)$, where the vertex function  $F(\bi{p},\bi{p'},\varepsilon)$ is given by the integral equation
$$
F (\bi{p},\bi{p'},\varepsilon) = \Sigma(\bi{p},\bi{p'},\varepsilon) + \int \Sigma(\bi{p},\bi{k},\varepsilon) G_0(\bi{k},\varepsilon) F(\bi{k},\bi{p'},\varepsilon) d \bi{k}.
$$
Here $\bi{p}$,$\bi{p'}$ are the electron momenta before and after the electron-phonon interaction in a local region. The expression for $\Sigma(\bi{p},\bi{p'},\varepsilon)$ is analogous to (\ref{Xsi}) with the localization functions depending on the differences $\bi{p}-\bi{k}$ and $\bi{k}-\bi{p'}$.
As is well known, it is the vertex function that is related to the scattering amplitude $f=-mF/(2 \pi)$ which plays a major part in the theory of transport processes. The poles of Green's function define the energy spectrum of a quasi-particle, and the same applies to the $f=-mF/(2 \pi)$ function, which is a correction to Green's function. Consider now the most interesting and intriguing case which is realized in the system under study.  

As already pointed out in Section  \ref{sec:effpotential}, a bound "phonon'' state exists in the local electron-phonon interaction region, but in the 1D case it exists always, and in the 3D case - only for sizes above the critical level; therefore, the amplitude of scattering from state $p$ into state $p$ should have a pole \cite{Mahan}. In other words, one can write
\begin{equation}
\label{f}
f(p)=- \frac{1+i \delta}{\kappa+i p}
\end{equation}
where $\kappa$,  is the wave vector corresponding to the "phonon'' state, and  $\delta$ is the correction to the scattering amplitude deriving from the scattering being inelastic. The magnitude of  $\delta$ is proportional to $\Sigma(\varepsilon,p)$  (see (\ref{endIm})). For small electron momenta,  $\delta \sim (pa)^2$. Significantly, it is in this feature that scattering from impurities and phonons differs radically from that from a local region. Indeed, as shown convincingly above, in the case of spatially uniform interaction $\delta$ features a threshold behavior.
For small momenta, the denominator of Eq. (\ref{f})  becomes small, with the ensuing strong growth of the scattering amplitude resulting from the resonance effect. For certain dimensions of the region of local electron-phonon interaction, the electron will undergo not only elastic resonant scattering from such regions but strong inelastic scattering as well.

\section{Results}
Consider now the effects occurring only in the presence of local electron-phonon interaction. There are two reasons for these effects to become manifest. The first of them can be traced to the possibility of inelastic electron scattering from phonons under conditions favoring local electron-phonon interaction. For each value of electron momentum, there is an optimum size of the local region  $a_{max}$, at which the inelastic scattering is maximal. The second reason lies in that the creation of a bound state in a region with a size on the order of $a_c$ may bring about enhancement of both the elastic and inelastic scattering.
We start with expressing the elastic, $\sigma_{elastic}$, and inelastic, $\sigma_{inelastic}$, scattering cross sections through the scattering amplitude $f$ , which can be done similar to the case of electron scattering from an impurity \cite{cross}. 
\begin{equation}
\sigma_{elastic} = \frac{\pi}{p^2} |1-S|^2 
\end{equation}
\begin{equation}
\sigma_{inelastic} = \frac{\pi}{p^2} (1-|S|^2)
\end{equation}
where, $S=1+2ipf$. 
For elastic scattering, we come to the relation describing resonant electron scattering from a bound state:
\begin{equation}
\label{elas}
\sigma_{elastic}=\frac{4\pi}{\kappa^2+p^2} 
\end{equation}
Examining Eq. (\ref{elas}), we see that elastic scattering from a region with local electron-phonon interaction occurs in the same way as it does from an impurity whose potential is governed by the polaron binding energy and the size corresponding to the characteristic size of the localization region.
Scattering from a local region differs substantially in the dependence of the inelastic scattering cross section on the electron momentum. As already pointed out (see Section \ref{sec:imagSE}), the scattering with not in this case have a characteristic threshold. We finally obtain that the cross section of inelastic electron scattering from a local region with a bound state depends on momentum as
\begin{equation}
\label{cross_inelastic}
\sigma_{inelastic}= \pi \frac{\delta(p)}{p} \frac{2\kappa+p \delta(p) }{\kappa^2+p^2} 
\end{equation}
In the three-dimensional case with existing bound state, this  transforms for  $pa \to 0$ to
\begin{equation}
\sigma_{inelastic}=4 \pi a^2 \frac{p}{\kappa}
\end{equation}
The inelastic scattering cross section can be made arbitrarily large provided the bound state energy is low enough, and, hence,  $\kappa \to 0$.

Figure 2 plots the inelastic scattering cross section as a function of momentum for the one-dimensional case in which the bound state always exists. In principle, the dependence of $\sigma_{inelastic}$ on momentum reproduces on a qualitative footing the dependence of the imaginary part of SE on momentum. We may recall, however, that this system is characterized by two essentially different parameters, $a_{max}$ and $a_c$. These sizes will become manifest in the dependence of the inelastic scattering cross section on size in the presence of two maxima. Although the inequality $a_c \le a_{max} \approx a_0$ for the characteristic parameters does hold, the graphs, in order to make them more revealing, were constructed in both cases both for $a_c \le a_{max}$  and for $a_c \ge a_{max}$.

\begin{figure}
\includegraphics[width=0.9\linewidth]{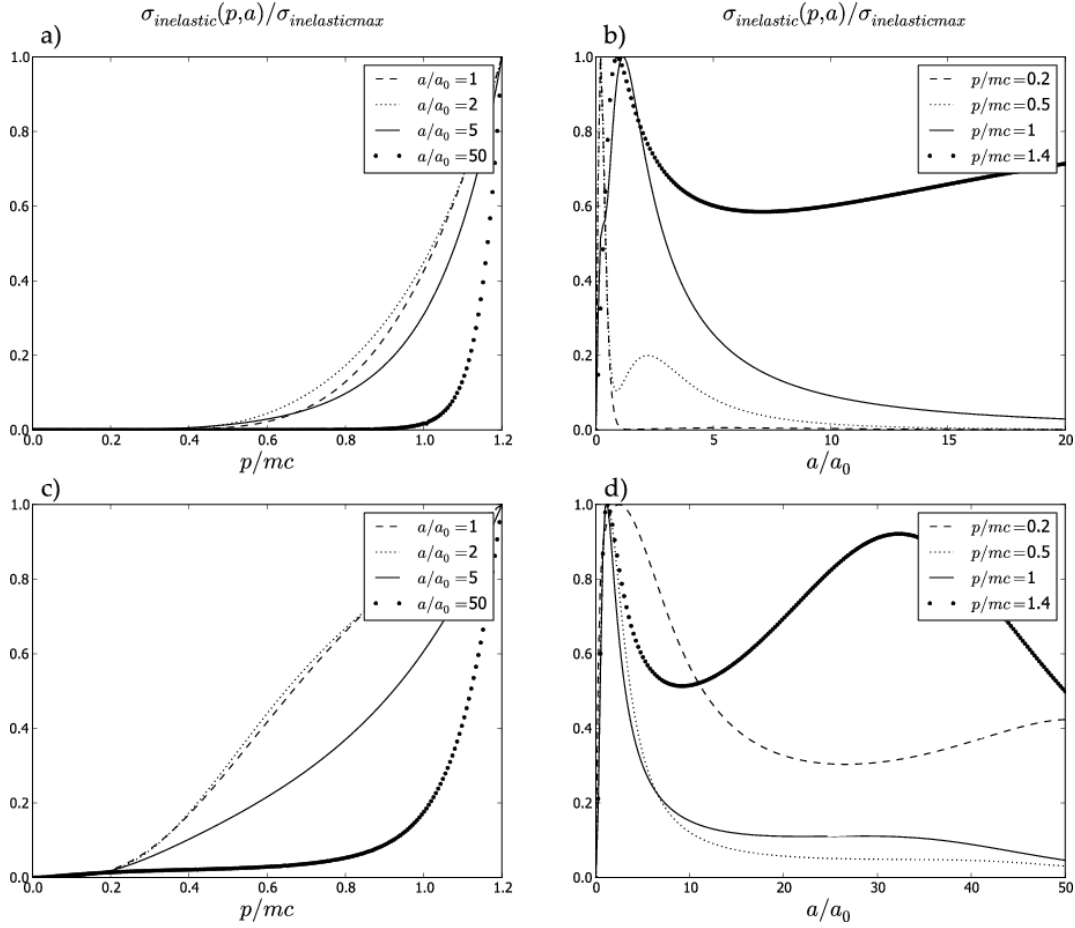}
\caption{
Graphs of the inelastic scattering cross section  $\sigma_{inelastic}$ for the parameter ratios $a_c/a_0=1/10$ ((a) and (b)) and $a_c/a_0=10$ ((c) and (d)). (a) and (c) plot the dependence on momentum for a given size, and (b) and (d), that on size for the specified momenta. As seen from (a) and (c), in the limiting cases of large sizes of the region, which correspond to uniform electron-phonon interaction, inelastic scattering appears only for momenta $p > mc$. For small local region sizes, the inelastic scattering cross section is nonzero for all, including small, values of the momentum. As evident from (b) and (d), for any momentum of the incident electron the inelastic scattering cross section features two maxima. These maxima  can be identified with the characteristic sizes of the electron-phonon interaction region, $a_c$ and $a_{max}$. The momentum $p$ and the size $a$ are presented in the same units as in Fig. \ref{fig1}. To present all the graphs on the same scale, they are normalized against their maximum values  $\sigma_{inelastic~max}$ within a given argument interval.}
\label{fig2}
\end{figure}

An important parameter in mesoscopic systems is the electron dephasing time, i.e., coherence loss time. At low temperatures, the dephasing time is mediated by various effects, among them the Kondo effect, scattering from impurities, and electron-electron interaction. It has recently become clear that scattering at low temperatures can likewise be initiated by interaction of electrons with phonons in local regions.
A mechanism of such interaction has been proposed for materials with clearly pronounced metallic properties  \cite{Dora}. In most semimetals, as well as in graphene and nanotubes, the electron gas is non-degenerate, which makes the mechanism forwarded in Ref.  \cite{Dora} inoperative. We have just shown (see Eq. (\ref{cross_inelastic})) that taking into account the finite size of the electron-phonon interaction region also gives rise to the appearance at low temperatures of a contribution to the dephasing time $\tau_{ph}$. 

In the three-dimensional case and for low momenta, the temperature dependence, according to Eqs. (\ref{cross_inelastic}, \ref{endIm}), should follow the inverse pattern, $\tau_{ph} = 1/ n p \sigma_{inelastic} \sim 1/T$.

This effect should be the strongest in one- and two-dimensional systems. The existence of a bound level in such systems should give rise to resonance effects, for any size of the electron-phonon interaction localization region (see Eq. (\ref{f})). Inelastic scattering, in particular, should appear in diamond-decorated nanotubes \cite{Reich1}. An electron propagates through a nanotube along its axis, and regions featuring local electron-phonon interaction are actually regions where diamond is in contact with the nanotube. In this case, the dephasing time should likewise scale as the inverse of the inelastic scattering cross section, $\tau_{ph} \sim 1/ p \sigma_{inelastic}$, and the dependence of the inelastic scattering cross section on electron  momenta and the size of decorated diamonds will be essentially nonmonotonic, as seen from Fig. \ref{fig2}

Graphene is a typical example of a two-dimensional system. The as-formed graphene can interact with the substrate. For an interaction region with the size favoring appearance of a bound state $a_c$ in a two-dimensional system, strong inelastic scattering can set in in this case as well. We believe that it is this situation that accounts for the observation of strong inelastic scattering reported recently \cite{Giant}.

The above analysis reveals another interesting ramification bearing on heat-conducting systems. We have been witnessing recently a broad research effort in the area of metal-insulator composites \cite{Kidalov}. We note, however, that in a metal it is electrons that transfer heat, while in insulators these are phonons. Therefore, the conduction of heat in such composites is mediated by the effects involved in energy transfer from the electron to the phonon subsystem. The major contribution comes from the processes taking place in the surface layer. It can readily be shown that Kapitza resistance coefficient \cite{Kapitza} $\chi$ at the metal-insulator interface should be inversely proportional to the cross section of inelastic scattering, $\chi \sim 1/ \sigma_{inelastic}$, of electrons from phonons in such a near-surface region. The above consideration fully applies here. For surface-region thicknesses on the order of $a_c$ or $a_{max}$, thermal resistance at the metal-insulator boundary should be substantially smaller than that in regions with other dimensions. This should be assigned to the strong inelastic scattering occurring in such a region (see Eq. (\ref{f})). Thus, to attain effective heat transport in metal-insulator composites, one should provide conditions favoring creation of a near-surface region where the above effects would become realized. The choice of the specific size depends on temperature.

\section{Conclusion}

The paper reports on a solution to the problem of electron interaction with phonons within a local region, which is based essentially on the modified Fr\"{o}hlich's Hamiltonian. It is demonstrated that such a system can support existence of a bound electron state within a local electron-phonon interaction region. The existence of such a bound state, just as that of a bound state in an impurity, translates into a substantial increase of the cross section of elastic scattering from such a region.

While the probability of phonon emission by an electron in a spatially uniform case follows a threshold pattern, in the system under consideration the probability of phonon emission by an electron, even by a slow one, is nonzero. This probability of phonon emission by an electron will become still higher in the presence of a bound state in the region of local electron-phonon interaction, thus initiating strong inelastic scattering. Therefore, local electron-phonon interaction will affect strongly the dephasing time in such systems.

\ack 
The author would like to thank E.D. Eidelman, A.Ya.Vul' for their time and advice. The author acknowledge support of the Dynasty Foundation and of the Federal Program "Research and Teacher Communities for Innovation in Russia, 2009-2013'' (State Contract $\Pi$182).

\end{document}